\documentclass[conference]{IEEEtran}
\IEEEoverridecommandlockouts
\usepackage{cite}
\usepackage{amsmath,amssymb,amsfonts}
\usepackage{algorithmic}
\usepackage{graphicx}
\usepackage{textcomp}
\usepackage{xcolor}
\usepackage{balance}
\usepackage{comment}
\usepackage{xcolor}
\usepackage{caption}
\usepackage{subcaption}
\usepackage{enumitem}
\usepackage{tikz}
\def\BibTeX{{\rm B\kern-.05em{\sc i\kern-.025em b}\kern-.08em
    T\kern-.1667em\lower.7ex\hbox{E}\kern-.125emX}}
\begin{document}
\bstctlcite{IEEEexample:BSTcontrol}

\title{Empirical Application Insights on Industrial Data and Service Aspects of Digital Twin Networks
}

\author{
\IEEEauthorblockN{ 
Marco Becattini\IEEEauthorrefmark{1},  
Davide Borsatti\IEEEauthorrefmark{2},
Armir Bujari\IEEEauthorrefmark{2}, 
Laura Carnevali\IEEEauthorrefmark{1},
\\
Andrea Garbugli\IEEEauthorrefmark{2}, 
Hrant Khachatrian\IEEEauthorrefmark{3}
Theofanis P. Raptis\IEEEauthorrefmark{4}, 
Daniele Tarchi\IEEEauthorrefmark{2}
}
\IEEEauthorblockA{
\IEEEauthorblockA{\IEEEauthorrefmark{1}University of Florence, Italy. Email: \{marco.becattini, laura.carnevali\}@unifi.it}
\IEEEauthorblockA{\IEEEauthorrefmark{2}University of Bologna, Italy. Email: \{davide.borsatti, armir.bujari, andrea.garbugli, daniele.tarchi\}@unibo.it}
\IEEEauthorblockA{\IEEEauthorrefmark{3}Yerevan State University, Armenia. Email: hrant.khachatrian@ysu.am}
\IEEEauthorblockA{\IEEEauthorrefmark{4}National Research Council, Italy. Email: theofanis.raptis@iit.cnr.it}
}
\vspace{-1cm}
}

\maketitle

\begin{tikzpicture}[remember picture,overlay]
\node[anchor=south,yshift=10pt] at (current page.south) {\fbox{\parbox{\dimexpr\textwidth-\fboxsep-\fboxrule\relax}{
  \footnotesize{
     \copyright 2024 IEEE. Personal use of this material is permitted.  Permission from IEEE must be obtained for all other uses, in any current or future media, including reprinting/republishing this material for advertising or promotional purposes, creating new collective works, for resale or redistribution to servers or lists, or reuse of any copyrighted component of this work in other works.
  }
}}};
\end{tikzpicture}

\begin{abstract}
Digital twin networks (DTNs) serve as an emerging facilitator in the industrial networking sector, enabling the management of new classes of services, which require tailored support for improved resource utilization, low latencies and accurate data fidelity. In this paper, we explore the intersection between theoretical recommendations and practical implications of applying DTNs to industrial networked environments, sharing empirical findings and lessons learned from our ongoing work. To this end, we first provide experimental examples from selected aspects of data representations and fidelity, mixed-criticality workload support, and application-driven services. Then, we introduce an architectural framework for DTNs, exposing a more practical extension of existing standards; notably the ITU-T Y.3090 (2022) recommendation. Specifically, we explore and discuss the dual nature of DTNs, meant as a digital twin of the network and a network of digital twins, allowing the co-existence of both paradigms. 
\end{abstract}


\section{Introduction}

Digital twins (DTs) represent virtual replicas of physical objects, systems, or processes, providing a mean to simulate, analyze, and optimize their behavior. 
The emergence of Digital Twin Networks (DTNs) marks a significant advancement in this paradigm. Unlike traditional DTs, which operate in isolation, DTNs 
also embody the concept of a ``DT of the network''. In this capacity, the DTN itself represents a virtual counterpart of the entire industrial network environment, providing real-time insights, monitoring capabilities, and predictive analytics on the network performance, health, and behavior. By encapsulating the collective intelligence and operational (real or synthetic) data of the interconnected digital twins, the DTN serves as a holistic representation of the industrial ecosystem, offering a comprehensive understanding of the network dynamics, vulnerabilities, and opportunities for optimization. Moreover, as a DT of the network, the DTN facilitates adaptive decision-making, enabling stakeholders to proactively respond to changing conditions, mitigate risks, and maximize the efficiency and resilience of the overall network infrastructure. 

DTNs have shown to be an effective tool to enable digitalization of complex ecosystems and effective analysis related to the mutual relationships between elements belonging to different domains within a common scenario\cite{jiang2021industrial}, specifically, in this study, an industrial setting. However, creating a DTN that encompasses data, architecture, and services is made difficult by historically different perspectives on how to digitalize and analyze those domains. DTs are born to represent a specific asset, namely industrial asset, and to track therefore their lifecycle. 
Conversely, DTs, once evolved from their original association to physical assets, have transformed to be able to represent also digital assets and hybrid elements, called phygital, which have both a physical and digital nature. Such elements are not by themselves a network, but their mutual relationship can be modelled and represented as a network indeed. This allows, for example, to enable hierarchization, collection, synthesis mechanics which are useful in digital domains \cite{elbazi2023generic}. Therefore, the concept of DTN as a network of DTs has also emerged. The two approaches, the DT of a network and the network of related DTs have so far coexisted, but the research on how they can be juxtaposed is still open~\cite{wieme2024managing}.

Data representations and fidelity, mixed-criticality workloads, and application-driven services play a pivotal role in shaping the effectiveness and applicability of DTNs within industrial environments \cite{9429703}. Firstly, the accuracy and fidelity of data representations are paramount for ensuring the reliability and validity of insights derived from DTNs. Inaccurate or incomplete data can compromise the integrity of decision-making processes and hinder the effectiveness of predictive analytics and optimization algorithms. Secondly, the presence of mixed-criticality workloads, encompassing tasks with varying levels of importance and urgency, poses unique challenges for DTNs, as they must accommodate diverse performance requirements while maintaining overall system stability and reliability. Thirdly, the ability of DTNs to support application-driven services is essential for meeting the specific needs and objectives of industrial stakeholders, ranging from real-time monitoring and control to advanced analytics and optimization.

Moreover, it is worth noting that the depth of detail of such aspects is often beyond the scope of conceptual DTN architectures, such as the ITU-T Y.3090 (2022) recommendation. While such standardized architectures provide valuable guidelines for the design and implementation of DTNs, they can potentially lack the granularity and specificity required to capture the intricacies of real-world industrial scenarios. Consequently, there is a need to complement conceptual architectures with empirical insights and practical considerations, derived from real-world experimentation and deployment of DTNs. By bridging the gap between theory and practice, and by incorporating insights from empirical studies, we can enhance the relevance, robustness, and effectiveness of DTNs in addressing the evolving needs and challenges of industrial networking environments.

\begin{figure}[t!]
    \centering
    \includegraphics[width=\columnwidth]{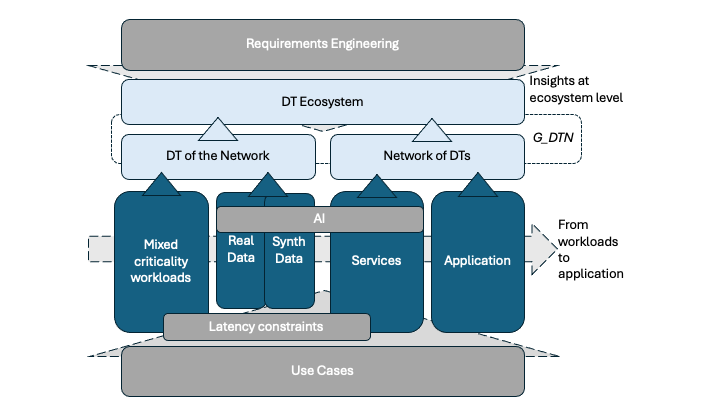}
    \caption{Logical diagram of our approach.}
    \label{fig:high}
    \vspace{-0.5cm}
\end{figure}

\emph{Our contribution}: In this paper, we explore the intersection between theoretical recommendations and practical implications in DTNs. Although organizations such as the International Telecommunication Union (ITU) and the Internet Research Task Force (IRTF) have established recommendations \cite{Y.3090} and work items \cite{irtf-nmrg-network-digital-twin-arch-04} on DTN requirements and conceptual architectures, there is still a lack of exploration into the practical aspects and implications of deploying DTNs, given the aforementioned challenges. Drawing from our working experience in networking, data management, service provisioning, and application deployment, we aim at providing valuable insights regarding selected aspects of DTNs.  Fig. \ref{fig:high}, displays the logical diagram of our aproach.

The roadmap of the paper features insights on concrete examples and is as follows: In Section \ref{ref:data}, we explore data representation and fidelity aspects, focusing on the utility of synthetic datasets for the industrial use case of environment reconstruction in wirelessly networked environments. In Section \ref{sec:crit}, we underscore the importance of virtualized workloads in industrial networking, highlighting challenges in maintaining quality of service (QoS) across diverse industrial applications. In Section \ref{sec:app}, leveraging platforms like Kubernetes and techniques like Mixture Density Networks, we demonstrate how DTNs can accurately model statistical distributions of microservice response times. Then, in Section \ref{sec:dt}, we introduce our concept of Generalized Digital Twin Networks aiming at managing and analyzing complex ecosystems in Industry 4.0 settings, bridging properties between network DTs and interconnected DTNs, with a focus on hierarchizability, abstraction, synthesis, modeling, and instantiability. Finally, in Section \ref{sec:open}, we outline the open challenges of this work.



\section{Data representations, fidelity and prediction} \label{ref:data}


To allow the creation and support of networking services that could nurture and effectively integrate industrial assets, it is relevant that the DTN has a clear and detailed vision of the status and capabilities of the network \cite{9440709}. Such vision can be obtained mainly with two strategies, which are not self-exclusive and can be combined for enhanced results:
\begin{enumerate}
    \item Having detailed replicas of the apparatus of the network, whereas here with apparatus we define every asset which comprises the network and with which the network should interface; such replicas are usually defined in DT domain as Doppelgänger \cite{9043088}, or simply Doppel, due to their near one-to-one virtualized replica of the real asset;

    \item Having a stylized, simplified models approach, based on capturing the essential information that is relevant to build a model and to allow full control on it, albeit forsaking the objective of a one-of-a-kind reconstruction of the original asset, preferring a more compact and functional stylized one. Such DTs are usually called within the domain Lightweight \cite{9931961}, or simply light, due to their more objective driven nature, which refute to provide a faithful descriptive representation of the original asset.
\end{enumerate}
Obtaining cyber-physical data, essential for creating detailed replicas or stylized models within DTNs, can often be challenging due to various factors such as accessibility, privacy concerns, and the sheer complexity of modern network infrastructures. However, synthetic dataset generation offers a promising alternative to complement or even overcome these necessities. By generating synthetic datasets that mimic real-world scenarios, use case owners can create a diverse range of data samples for training and testing DTN models. For instance, recent datasets such as WAIR-D \cite{huangfu2022waird} and RadioUnet \cite{10122907} provide rich sources of synthetic data for wireless communication networks, enabling us to validate DTN models without relying solely on real-world data. This approach not only addresses data accessibility issues but also allows for controlled experimentation and validation of DTN concepts and algorithms.

Deep learning (DL) techniques on such synthetic datasets can offer significant potential for enhancing the capabilities of DTNs, particularly in use cases involving low-energy, low-compute devices such as augmented reality navigation systems. By leveraging DL algorithms, DTNs can effectively process and analyze complex data streams to reconstruct environments, optimize resource allocation, and support decision-making processes in real-time  \cite{10233994}. For example, DL models can be trained to reconstruct detailed environmental maps from just radio frequency (RF) data captured by networked devices deployed in environments such as industrial ones. These reconstructed maps can then be used to improve navigation accuracy and efficiency in DTN applications, where low-latency responses are critical to the system efficiency.

\begin{figure}[tbp]
     \centering
     \begin{subfigure}[b]{0.19\textwidth}
         \centering    \includegraphics[width=\textwidth]{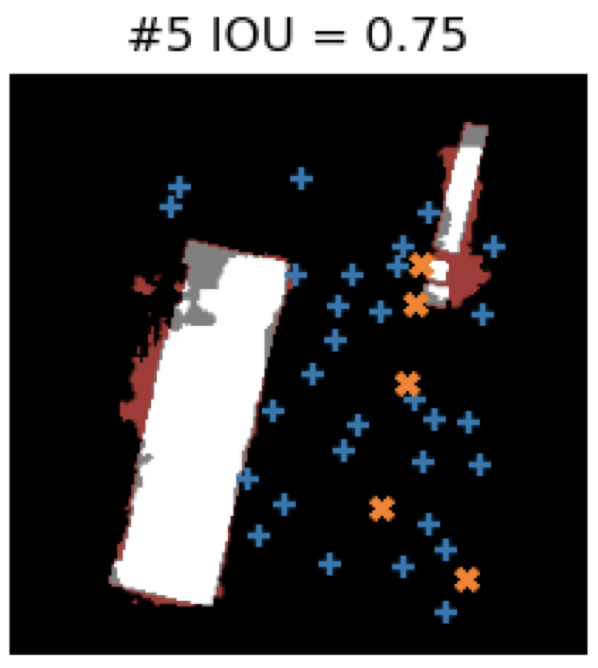}
         \caption{High fidelity.}
         \label{fig:iou1}
     \end{subfigure}
     \hfill
     \begin{subfigure}[b]{0.19\textwidth}
         \centering
         \includegraphics[width=\textwidth]{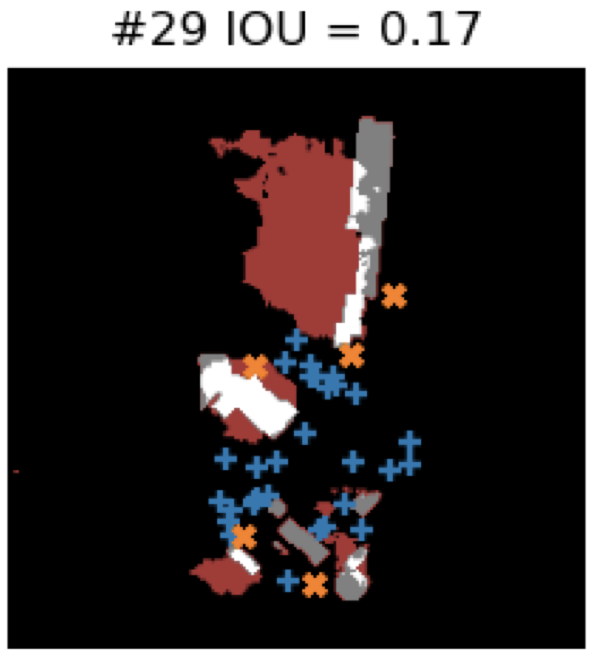}
         \caption{Low fidelity.}
         \label{fig:iou2}
     \end{subfigure}
    \caption{Environment digital reconstruction.}
    \label{fig:iou}
    \vspace{-0.5cm}
\end{figure}

In our experimentation with the WAIR-D dataset \cite{huangfu2022waird}, we used deep learning techniques to digitally reconstruct environments from wireless RF signals \cite{khachatrian2024outdoor}. Figure \ref{fig:iou} depicts two indicative results of the environment reconstruction achieved using a transformer-based architecture trained on the WAIR-D dataset, for two different urban environments. We measure performance using the intersection-over-union metric (IoU) \cite{iou} and the visual representation of the digitally reconstructed result. White and gray areas correspond to buildings, whereas gray areas correspond to parts of the buildings that are not detected by our method (false negatives). Dark red areas correspond to false positives. Orange and blue crosses indicate the locations of wireless base stations and wireless user equipment, respectively.

In Figure \ref{fig:iou1}, the reconstruction exhibits a notably high Intersection over Union (IoU) accuracy, indicating a close match with the ground truth environment. The success of this reconstruction can be correlated with the spatial distribution of nodes that capture input signals, which significantly influences the accuracy of the resulting digital representation. Figure \ref{fig:iou2} showcases another instance of environment reconstruction from a different perspective. While maintaining the same training dataset, the reconstruction in Figure \ref{fig:iou2} demonstrates lower effectiveness in capturing intricate environmental features. These results underscore the importance of input data distribution in achieving accurate digital reconstructions within DTNs, highlighting the potential for enhanced spatial awareness and application versatility.

\section{Layered management and orchestration for mixed-criticality workloads} \label{sec:crit}

Pushing towards the introduction of virtualized workloads, whereby industrial applications are packaged and deployed as disaggregated, virtualized software entities is another imperative in the industrial networking domain \cite{8764545}. This paradigm shift leverages cloud computing practices, relying on containers and orchestration systems such as Kubernetes for improved resource utilization due to multiplexing. However, provisioning this capability requires particular attention, especially when considering mixed-criticality industrial environments serving a wide range of (control) applications ranging from traditional, time-sensitive, and reliable closed-loop control systems to event-driven, delay-bounded, or best-effort sensor traffic. In this perspective, the main challenge is to provide methodologies, algorithms, and mechanisms for the joint QoS management of virtualized application resources and potentially heterogeneous networks, ensuring the end-to-end QoS of co-existing applications.

The above value-added proposition can be achieved if the solution aligns and considers: 

\begin{enumerate}

\item A continuum of resources in a holistic way, where resource management decisions are taken cross-layer by jointly considering computing/storage/networking at edge nodes, avoiding transitory phenomena which can manifest when considering a resource-centric, layered approach~\cite{bujari2023layered}.

\item Models and techniques to manage virtualized resources in a QoS-dependent and unified manner across different network segments and edge nodes. This requires a unifying semantic to define application QoS requirements and a set of models to map and align them into technology-dependent QoS requirements on computing, storage, and networking resources. This capability could pave the way for the adoption of the nascent Intent Based Networking approach, allowing industrial operators to express high-level goals/objectives~\cite{borsatti2022category}. 

\item Dynamicity as a central feature and requirement. Distributed resource management decisions can be statically taken at the beginning, but dynamically refined and reconsidered, in an autonomous and distributed way, every time i) new tasks/applications for the production line are introduced (admission control for reconfigurable I4.0 factories); ii) the best-effort and non-deterministic nature of some virtualized components in the control loop chain generates significant discrepancies from the planned QoS behavior~\cite{bellavista2021application}. 

\end{enumerate}


Indeed, in addition to the initial deployment and configuration actions, at runtime, there is the need to monitor the achieved end-to-end QoS and the possibly new applications dynamically deployed in the industrial plant to take proper management actions to prevent/mitigate QoS degradation. Actions might involve the migration of microservices between edge nodes. Techniques for efficient stateful/stateless live/cold migration of the workload can be exploited to implement re-allocation, also based on workload predictions, accommodating QoS fluctuations and/or reacting to failures. 

\begin{figure}[t!]
    \centering
    \includegraphics[width=0.8\columnwidth]{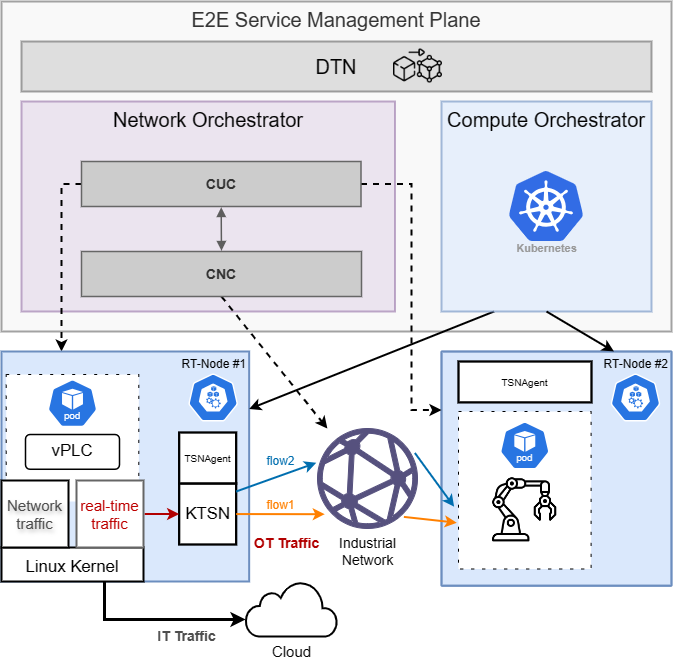}
    \caption{Layered architecture for end-to-end QoS management of mixed-critical workloads.}
    \label{fig:e2e-industrial}
    \vspace{-0.5cm}
\end{figure}

To this end, we display some insights gained from our development effort of a system composed of multiple independent yet interconnected modules, such as the Centralized Network Configuration (CNC), the Centralized User Configuration (CUC), a module supporting the DT of the managed network, and a Time Sensitive Network (TSN) agent. The architecture shown in Fig.~\ref{fig:e2e-industrial} depicts the interaction of the different entities, implementing a so-called centralized management \& control solution for Time Sensitive Networks (TSN)~\cite{qosTSN}. In this setting, the CNC communicates with network devices via specific protocols (e.g., NETCONF), managing QoS enforcement, synchronization, VLAN setting, and distribution of flow schedules in the network. It also receives event notifications for monitoring metrics. The CUC collaborates with the TSN Agent to manage the TSN flow configurations in end devices through a set of specific APIs. The DT stores information about the managed elements and is updated periodically through the CUC-CNC interface. The TSN Agent, deployed on the end devices, communicates with the CUC, receiving configuration parameters for setting up QoS-aware TSN flows and using operating system capabilities to configure and monitor the network devices involved in the communication.

\begin{figure}[t!]
    \centering
    \includegraphics[width=0.8\columnwidth]{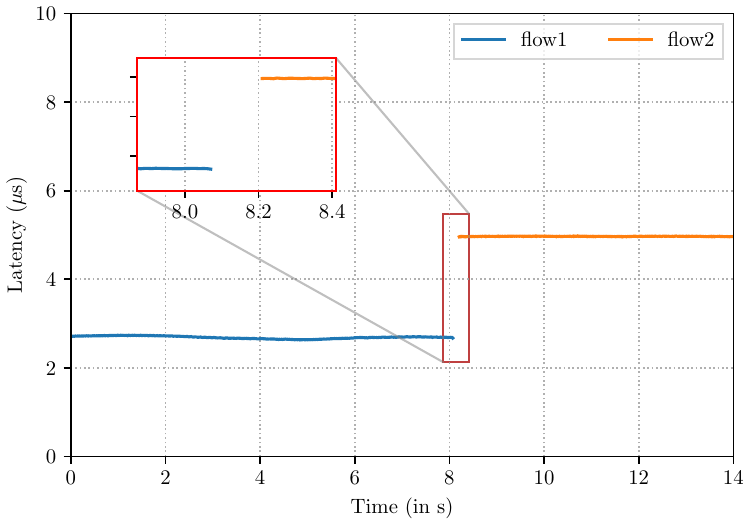}
    \caption{Observed latencies before and after the reconfiguration phase, passing from \emph{flow 1} to \emph{flow 2}.}
    \label{fig:reconf}
    \vspace{-0.5cm}
\end{figure}

To validate the system, we set up a physical testbed with two nodes that can communicate via two physical paths: \emph{flow 1} and \emph{flow 2}. In the initial phase, both the network and the devices are configured to communicate using flow 1. We then simulate the failure of a link in path 1. Thanks to the observability mechanisms implemented, the system can react autonomously and reconfigure both the network and the nodes to use the second available path and keep the communication active. In Fig.~\ref{fig:reconf}, we show the end-to-end latency, measured as the difference between the receive and transmit time at the kernel level, before and after the route change. The difference in terms of latency between flow 1 and flow 2 is given by the different number of network hops for the two different paths. Furthermore, due to the reconfiguration time of both the network and the end devices, we have a downtime of 150 ms.

\section{Application-driven services} \label{sec:app}

The creation of a DTN that can fully represent and allow fine-grain control on the overall ongoing functioning of the network, but especially on the elements deployed on the edge, that, as we have seen, provide unique characteristics in terms of latency, bandwidth, and computational power, would enable the creation of novel type of services.

Specifically, DTs offer the capability to have continuous real-time, or near-real-time, knowledge of the status of the network and services, both to the network provider and the users. This is key to allowing for the construction of services which make efficient use of low latency, high bandwidth, and fast and massive computational power on site.

Such an amount of awareness and control on the status of the industrial networks and related offered services allows the creation of new paradigms that transcend the merely SLA (Service Level Agreement) concept, which is restrictive and not suited to model, also in business and merely technical terms, the complexity of this new kind of services enabled by edge deployed capabilities and their needs, which are demanding and constantly time-changing, potentially both topologically and in terms of needs (latency, bandwidth, computational power) \cite{9696282}.

Such new paradigms can offer a solid foundation for creating a new class of services that can leverage the unprecedented level of service that industrial networks can offer and simultaneously handle the tight and high-level needs and constraints that services such as enhanced or virtual reality require.

In order to have a well-performing DT of these distributed environments, we need a strong time characterization of the workloads developed in it. With this knowledge, operators can test the performance of their system with varying loads and experiment with diverse workload distributions without interfering with the real system. We decided to focus on our study on Kubernetes, which is the de facto standard for the orchestration of container-based applications. Furthermore, Anuket, an initiative backed by the Linux Foundation, GSMA, and other telco providers, has identified Kubernetes as one of the two reference architectures for a industrial networking cloudification \cite{anuketRa2}.

To this end, we first developed a DT system for Kubernetes environments called KubeTwin, which can reenact a Kubernetes cluster's main procedures and functionalities \cite{10217853} (Kubernetes is relevant to industrial networks for its ability to efficiently manage, scale, and ensure the high availability of containerized applications while providing flexibility and security across diverse deployment environments). Then, we started investigating how to find good statistical distributions of the response time of microservices running inside the cluster. 
We used the new methodology based on Mixture Density Networks (MDNs), which we introduced in \cite{10327842}, to accurately estimate the statistical distribution of the response time of microservice-based applications. 
MDNs are a powerful type of neural network that tackles a shortcoming of standard ones. 

\begin{figure}[t!]
    \centering
    \includegraphics[width=0.8\columnwidth]{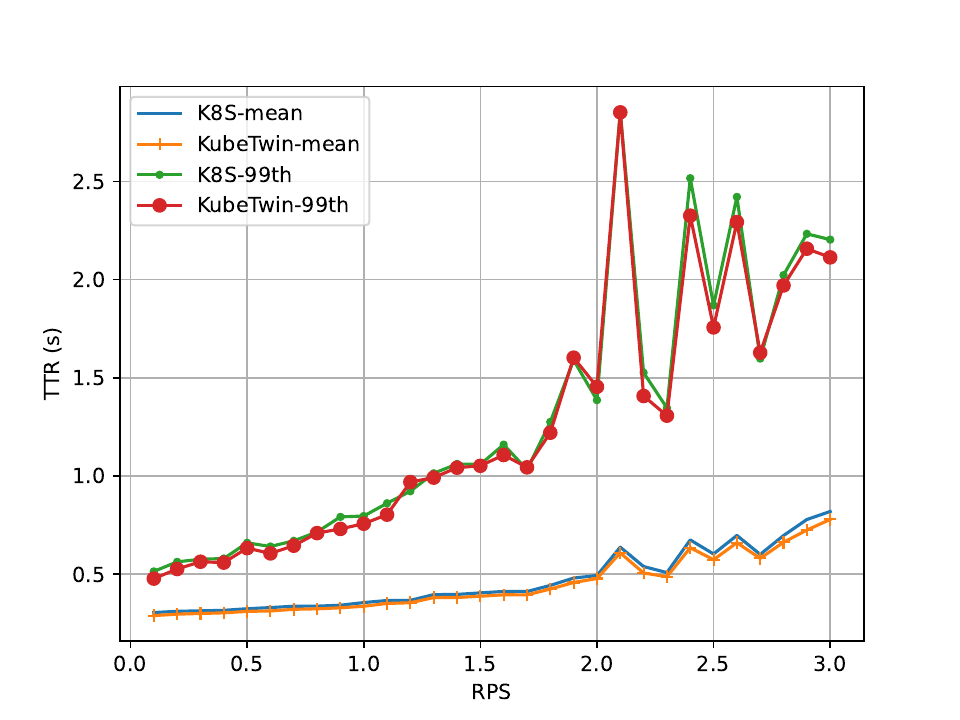}
    \caption{Comparison between the mean and 99th percentile response time of the real application and KubeTwin.}
    \label{fig:comp-mttr-99-20-rps}
    \vspace{-0.5cm}
\end{figure}

While regular neural networks give a single prediction, MDNs can capture the entire probability distribution of what they are predicting. In this way, we are not only predicting a microservice next response time value but also the statistical distribution of its response time. Therefore, these statistical distributions can be fed to KubeTwin, which can sample from them to reenact their behavior in its simulations closely. We trained these MDNs over real traces of an image recognition application composed of two microservices, obtaining their statistical distribution. We then fed this distribution to KubeTwin and observed that the DT could replicate the behavior of the real system with satisfactory accuracy, even under request loads unseen during training. An example of these results can be seen in Fig.~\ref{fig:comp-mttr-99-20-rps}. The figure presents a comparison between the response time under different traffic loads of a real run of the modeled application inside a Kubernetes cluster with the result obtained from the simulation in the DT. 

\section{Toward a practical DT network architecture} \label{sec:dt}




We worked towards a novel approach, by introducing the concept of  Generalized Digital Twin Networks (G\_DTNs), designed to oversee the lifecycle management and analysis of complex ecosystems encompassing both network DTs and interconnected DTNs. This approach is particularly suited for Industry 4.0 and beyond, modeling the complex ecosystems involving both the connectivity infrastructure and the web of assets, services, and applications governing a given industrial setting. Thanks to their inherent duality, G\_DTNs are suited to encapsulate the multifaceted ecosystems characteristic of advanced industrial paradigms, offering a robust platform for the digitalization and analytical scrutiny required in these complex environments. Our G\_DTN approach aims at providing the following properties to reduce the differences between a network of DT and a DTN:




 
 
 
 


\begin{itemize}

\item \emph{Hierarchizability:} this property enables developers to design and structure hierarchies of DTs, where lower-tier DTs more closely model the assets themselves, and upper-tier DTs provide broader network-related information. 

\item \emph{Abstraction:} this property allows DTs within a G\_DTN to adopt a stylized fact representation of both themselves and the subordinate DTs. It supports alternative representations that deviate from a traditional DT model, which typically seeks to provide an accurate replica of the underlying assets.

\item \emph{Synthesis:} without the ability to synthesize, abstraction could be largely ineffective. The synthesis property facilitates the creation of a meaningful and valuable stylized fact representation, abstracted from the collection of information across a hierarchy of DTs. This synthesized information provides a foundation for further investigation and insights.

\item \emph{Modeling:} by leveraging stochastic modeling, the virtualized representation of a DT can be transformed into a stochastic model  supporting the pursuit of insights via quantitative methods. 

\item \emph{Instantiability:} analysis to seek network insights and optimization may be  performed in settings ranging across the cloud to edge continuum. This strategic placement allows for various advantages, including eco-friendly computing and simultaneous analyses (useful in what-if scenarios). The selection of the instantiation location depends on the type of analysis required and its associated constraints. For tasks demanding high speed and low latency, edge computing is preferred, while the cloud is better suited for complex analyses that are less sensitive to latency. This allows for several benefits to be leveraged, notably including green computation and multiple analysis running in parallel (useful to evaluate what-if scenarios).

\end{itemize}

\begin{figure}[t!]
    \centering
    \includegraphics[width=\columnwidth]{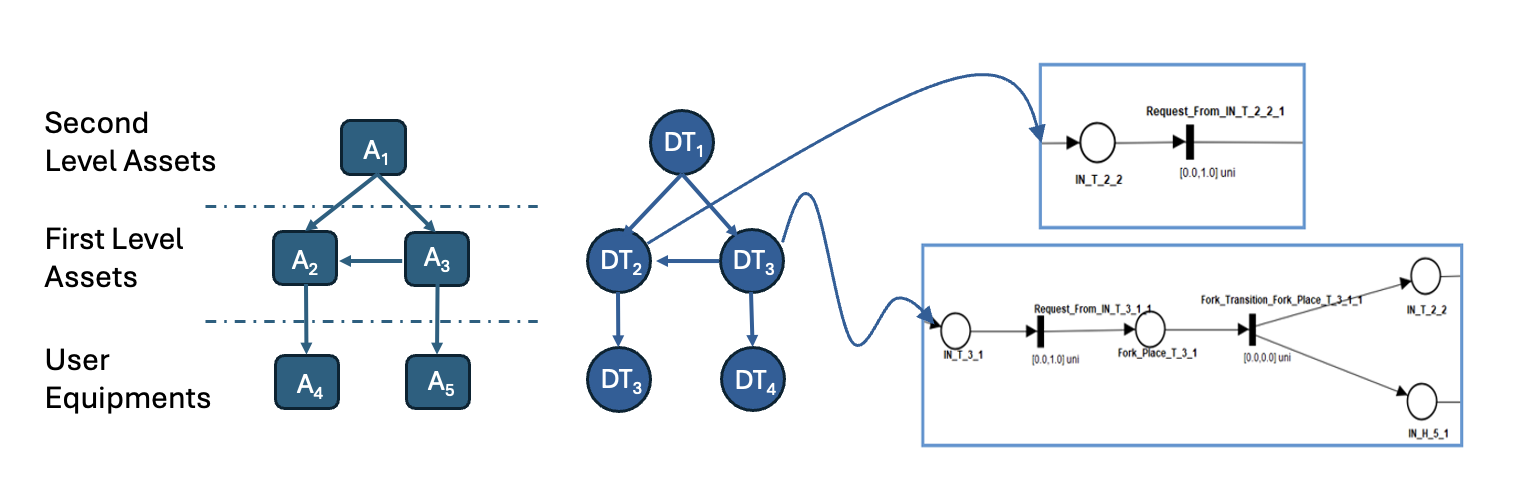}
    \caption{Construction of a G\_DTN, from the current assets to their digital representation to stochastic models.}
    \label{fig:G_DTN_process}
    \vspace{-0.5cm}
\end{figure}

To experimentally verify the above-outlined reasoning, a proof of concept (POC) has been created for G\_DTNs, following the priciples also of the reference architecture in ITU-T Y.3090, displayed in Fig. \ref{fig:G_DTN_process}. The POC is based on creating a model for DTs, implementing all the above-listed properties.  The POC has been verified to guarantee that it is able to model both the DT of a network and a network of DTs. To verify its capability of modeling a DT of a network, the architecture has been connected with a Virtual Network (VN), and simulated through the use of Docker network functionalities. Testing has been performed by connecting the POC to a Docker Virtual Network (DVN). In particular, we have verified each modeling step of the VN, from the association of digital assets to their respective DTs, to the automated creation of relationships between DTs in the DVN. Assets and their DTs have been associated with the edges of the network.

Preliminary results show that the POC is suitable to model a VN, creating a DT for each virtual asset within the VN. At the same time, the capability of modeling a network of DTs has been verified, by testing the capability of representing relationships between DTs, which can be part of the VN or elements external to the VN, such as industrial assets, but connected to the VN in any form (e.g., by a point-to-point connection). Consequently, it has been shown that an implementation of G\_DTNs is possible. 

Further experiments have shown that G\_DTNs can be simply transformed into a stochastic model, namely a Direct Acyclic Graph (DAG) with stochastic durations~\cite{carnevali2023compositional}. 
Note that DAGs naturally emerge as suitable models, given that a node can potentially have multiple parents, as in the case of an asset contributing to different sub-networks.
DAGs with stochastic durations can be further transformed into other stocastic models, e.g.,~Stochastic Time Petri Nets (STPNs) supported by the ORIS tool and its underlying SIRIO Java library, enabling evaluation of quantitative properties of interest.


\section{Open challenges} \label{sec:open}

The remaining challenges are the following:
\begin{itemize}
    \item As discussed in Section \ref{sec:dt}, experimental verification of the properties of G\_DTNs in a real environment still remains open. So far, experiments with have been fully conducted with in-vitro settings, using a virtual network.

    \item The use of G\_DTNs to gain insights on a complex environment remains open. So far, it has been shown that G\_DTNs can be transformed into stochastic models. The next step would be to investigate quantitative phenomena, such as evaluating the reliability of a complex system, specifically an industrial scenario.
    
    \item As highlighted in Section \ref{ref:data}, one of the key challenges in the advancement of DTN data representations and fidelity lies in achieving seamless interoperability between synthetic and real-world data sources. Although synthetic datasets offer scalability and control, ensuring or adapting their fidelity to real-world scenarios remains a significant challenge. This challenge is compounded by issues such as domain shifting and generalization limitations inherent in deep learning models \cite{10288574}. Furthermore, discrepancies between synthetic and real-world radio environments, known as radio mismatch, pose additional challenges in accurately modeling the wireless parts of DTNs. 
    

\end{itemize}

\section*{Acknowledgements}
Work partly supported by the European Union under the Italian National Recovery and Resilience Plan (NRRP) of NextGenerationEU, partnership on ``Telecommunications of the Future'' (PE00000001 - program ``RESTART''). The work of H. Khachatrian was partly supported by the RA Science Committee grant No. 22rl-052.

\balance
\bibliographystyle{IEEEtran}
\bibliography{refs}

\end{document}